\documentstyle[aps,floats,epsf,graphicx]{revtex}

\def\D{\mbox{D}}

\begin{document}

\twocolumn[\hsize\textwidth\columnwidth\hsize\csname
@twocolumnfalse\endcsname

\title{Stars in the braneworld}

\author{Cristiano Germani and Roy Maartens}

\address{~}
\address{Relativity and Cosmology Group, School of
Computer Science and Mathematics, Portsmouth University,
Portsmouth~PO1~2EG, Britain}

\maketitle

\begin{abstract}

We show that in a Randall-Sundrum~II type braneworld, the vacuum
exterior of a spherical star is not in general a Schwarzschild
spacetime, but has radiative-type stresses induced by
5-dimensional graviton effects. Standard matching conditions do
not lead to a unique exterior on the brane because of these
5-dimensional graviton effects. We find an exact uniform-density
stellar solution on the brane, and show that the general
relativity upper bound $GM/R<{4\over9}$ is reduced by
5-dimensional high-energy effects. The existence of neutron stars leads to a
constraint on the brane tension that is stronger than the big bang
nucleosynthesis constraint, but weaker than the Newton-law
experimental constraint. We present two different
non-Schwarzschild exteriors that match the uniform-density star on
the brane, and we give a uniqueness conjecture for the full
5-dimensional problem.

\end{abstract}

\vskip 1cm]

\section{Introduction}

String theory and M-theory describe gravity as a truly
higher-dimensional interaction, which becomes effectively
4-dimensional at low enough energies. In braneworld models
inspired by these theories, the observable universe is a 3-brane
(domain wall) to which standard-model fields are confined, while
gravity can access the extra spatial dimensions. Randall and
Sundrum~\cite{rs} showed how gravity could be localized near the
brane at low energies even with a noncompact extra dimension. The
warped spacetime metric satisfies the 5-dimensional Einstein
equations with negative cosmological constant. Their models have
been generalized to allow for arbitrary energy-momentum tensor on
the brane~\cite{sms}.

The cosmological implications of these braneworld models have been
extensively investigated (see e.g. the review~\cite{m2} for
further references). Significant deviations from Einstein's theory
occur at very high energies, as in the very early universe.
Gravitational collapse can also produce very high energies where
5-dimensional corrections would become significant. If an horizon
forms, then the high-energy effects eventually become disconnected
from the outside region on the brane. However, they could leave a
signature on the brane. In addition to local high-energy effects,
there are also nonlocal corrections arising from the imprint on
the brane of Weyl curvature in the bulk, i.e. from 5-dimensional
graviton stresses. These nonlocal Weyl stresses arise on the brane
whenever there is inhomogeneity in the density; the inhomogeneity
on the brane generates Weyl curvature in the bulk which
`backreacts' on the brane. Anyway we can have these nonlocal Weyl
stresses even if the density is homogeneous, as we show in the
case of static stars.

The high-energy (local) and bulk graviton stress (nonlocal)
effects combine to significantly alter the matching problem on the
brane, compared with the general relativistic case. For spherical
compact objects (uncharged and non-radiating), matching in general
relativity shows that the asymptotically flat exterior spacetime
is Schwarzschild. High-energy corrections to the pressure,
together with Weyl stresses from bulk gravitons, mean that on the
brane, matching no longer leads to a Schwarzschild exterior in
general. These stresses also mean that the matching conditions do
not have unique solution on the brane; knowledge of the
5-dimensional Weyl tensor is needed as a minimum condition for
uniqueness. In the non-static case, it seems likely that in
dynamic gravitational collapse to a black hole, the corrections to
the Schwarzschild exterior may also be non-static.

In this paper we consider the simplest case of a static spherical
star with uniform density. We find an exact interior solution,
thus generalizing the Schwarzschild interior solution of general
relativity. We show that the general relativity compactness limit
given by $GM/R<{4\over9}$ is reduced by high-energy 5-dimensional
gravity effects. The existence of neutron stars allows us to put a
lower bound on the brane tension, which is stronger than the bound
from big bang nucleosynthesis, but weaker than the bound from
experiments probing Newton's law on sub-millimetre scales. We also
give two different exact exterior solutions, both of which satisfy
the braneworld matching conditions and field equations and are
asymptotically Schwarzschild, but neither of which is the
Schwarzschild exterior. One of these solutions is the
Reissner-N\"ordstrom-type solution found in~\cite{dmpr}, in which
there is no electric charge, but instead a Weyl `charge' arising
from bulk graviton tidal effects. The other is a new solution.
Both of these exterior solutions carry the imprint of bulk
graviton stresses, and each has an horizon on the brane which is
larger than the Schwarzschild horizon.

 Both of our solutions (i.e. the full solution, interior plus exterior) are consistent
braneworld solutions, but we do not know the bulk solutions of
which they are boundaries. In fact, no exact 5-dimensional
solution for astrophysical brane black holes is known, and the
uniform star case is even more complicated. It is in principle
possible to integrate numerically into the bulk (assuming
appropriate boundary conditions) to find the 5-dimensional metric
for which these stellar solutions are brane boundaries. However,
even in the much simpler case of black hole solutions, numerical
integration into the bulk proves very difficult~\cite{crss}. In
the absence of exact or numerical solutions, further investigation
is needed into the 5-dimensional aspects of stellar solutions and
their exteriors. Perturbative studies of the static weak-field
regime~\cite{rs,pert,ssm} show that the leading order correction
to the Newtonian potential on the brane is given by
\begin{equation}\label{pert}
\Phi={GM\over r}\left(1+{2\ell^2\over3r^2}\right)\,,
\end{equation}
where $\ell$ is the curvature scale of 5-dimensional anti de
Sitter spacetime (AdS$_5$). This result assumes that the bulk
perturbations are bounded in conformally Minkowski coordinates,
and that the bulk is nearly AdS$_5$. It is not clear whether there
is a covariant way of uniquely characterizing these perturbative
results~\cite{dd}, and therefore it remains unclear what the
implications of the perturbative results are for very dense stars
on the brane. However, it seems reasonable to conjecture that the
bulk should be asymptotically AdS$_5$, and that its Cauchy horizon
should be regular. Then perturbative results suggest that on the
brane, the weak-field potential should behave as in
Eq.~(\ref{pert}). In fact, perturbative analysis also constrains
the weak-field behaviour of other metric components on the
brane~\cite{pert}, as well as of the nonlocal stresses on the
brane induced by the bulk Weyl tensor~\cite{ssm}.

\section{Field equations and matching conditions}

The local and nonlocal extra-dimensional modifications to
Einstein's equations on the brane may be consolidated into an
effective total energy-momentum tensor~\cite{sms}:
\begin{equation}
G_{\mu\nu}=\kappa^2 T^{\rm eff}_{\mu\nu}\,, \label{6'}
\end{equation}
where $\kappa^2=8\pi G$ and the bulk cosmological constant is
chosen so that the brane cosmological constant vanishes. The
effective total energy density, pressure, anisotropic stress and
energy flux for a perfect fluid are~\cite{m}
\begin{eqnarray}
\rho^{\rm eff} &=& \rho\left(1+{\rho\over 2\lambda}\right)
+{6\over \kappa^4\lambda}{\cal U} \label{a'}\\ p^{\rm eff} &=& p+
{\rho\over 2\lambda}\left(\rho+2p\right) +{2\over
\kappa^4\lambda}{\cal U} \label{b'}\\ \pi^{\rm eff}_{\mu\nu} &=&
{6\over \kappa^4\lambda}{\cal P}_{\mu\nu}\label{c'}\\ q^{\rm
eff}_\mu &=&{6\over \kappa^4\lambda}{\cal Q}_\mu \,,\label{d'}
\end{eqnarray}
where $\lambda$ is the brane tension, and general relativity is
regained in the limit $\lambda^{-1}\to0$.

From big bang nucleosynthesis constraints, $\lambda \gtrsim
1$~MeV$^4$, but a much stronger bound arises from null results of
sub-millimetre tests of Newton's law: $\lambda \gtrsim
10^8$~GeV$^4$~\cite{mwbh}.

The local effects of the bulk, arising from the brane extrinsic
curvature, are encoded in the quadratic terms, $\sim
(T_{\mu\nu})^2/\lambda$, which are significant at high energies,
$\rho\gtrsim\lambda$. The nonlocal bulk effects, arising from the
bulk Weyl tensor, are carried by nonlocal energy density ${\cal
U}$, nonlocal energy flux ${\cal Q}_\mu$ and nonlocal anisotropic
stress ${\cal P}_{\mu\nu}$. Five-dimensional graviton stresses are
imprinted on the brane via these nonlocal Weyl terms.

Static spherical symmetry implies ${\cal Q}_\mu=0$ and
\begin{equation}\label{p}
{\cal P}_{\mu\nu}={\cal P}(r_\mu r_\nu-{\textstyle{1\over3}}
h_{\mu\nu})\,,
\end{equation}
where $r_\mu$ is a unit radial vector, and
$h_{\mu\nu}=g_{\mu\nu}+u_\mu u_\nu$ projects into the rest space
of static observers with 4-velocity $u^\mu$. The brane
energy-momentum tensor separately satisfies the usual conservation
equations, $\nabla^\nu T_{\mu\nu}=0 $, and the 4-dimensional
Bianchi identities on the brane imply that the effective
energy-momentum tensor is also conserved: $\nabla^\nu T^{\rm
eff}_{\mu\nu}=0 $. For static spherical symmetry, these
conservation equations~\cite{m} reduce to
\begin{eqnarray}
&&\D_\mu p+(\rho+p)A_\mu=0\,,\label{c1}\\ && {\textstyle{1\over3}}
\D_\mu{\cal U}+ {\textstyle{4\over3}}{\cal U}A_\mu+\D^\nu {\cal
P}_{\mu\nu}= -{\textstyle{1\over6}}\kappa^4(\rho+p) \D_\mu\rho\,,
\label{c2}
\end{eqnarray}
where $\D_\mu$ is the covariant spatial derivative and $A_\mu$ is
the 4-acceleration. In static coordinates the metric is
\begin{eqnarray}\label{m}
 ds^2=-A^2(r)dt^2+B^2(r)dr^2+r^2d\Omega^2\,,
\end{eqnarray}
and Eqs.~(\ref{6'})--(\ref{m}) imply
\begin{eqnarray}
&&{1\over r^2}-{1\over B^2}\left({1\over r^2}-{2\over r}{B'\over
B}\right) = 8\pi G\rho^{\rm eff}\,,\label{f1}\\ &&-{1\over
r^2}+{1\over B^2}\left({1\over r^2}+{2\over r}{A'\over A}\right) =
8\pi G\left(p^{\rm eff}+ {4\over\kappa^4\lambda}{\cal P}
\right)\,,\label{f2}\\&&p'+{A'\over A}(\rho+p)=0\,,\label{f3}\\
&&{\cal U}'+4{A'\over A} {\cal U}+ 2{\cal P}'+2{A'\over A}{\cal
P}+ {6\over r}{\cal P}\nonumber\\ &&~~~~~~{}=-2(4\pi
G)^2(\rho+p)\rho'\,. \label{f4}
\end{eqnarray}

The exterior is characterized by
\begin{equation}
\rho=0=p\,,~~{\cal U}={\cal U}^+\,,~{\cal P}={\cal P}^+\,,
\end{equation}
so that in general $\rho^{\rm eff}$ and $p^{\rm eff}$ are nonzero
in the exterior: there are in general Weyl stresses in the
exterior, induced by bulk graviton effects. These stresses are
radiative, since their energy-momentum tensor is traceless
($p^{\rm eff}={1\over3} \rho^{\rm eff}$). The system of equations
for the exterior is not closed until a further condition is given
on ${\cal U}^+$, ${\cal P}^+$ (e.g., we could impose ${\cal
P}^+=0$ to close the system). In other words, from a brane
observer's perspective, there are many possible static spherical
exterior metrics, including the simplest case of Schwarzschild
(${\cal U}^+=0={\cal P}^+$).

The interior has nonzero $\rho$ and $p$; in general, ${\cal U}^-$
and ${\cal P}^-$ are also nonzero, since by Eq.~(\ref{f4}), {\em
density gradients are a source for Weyl stresses in the interior}.
For a uniform density star, we can have ${\cal U}^-=0={\cal P}^-$,
but nonzero ${\cal U}^-$ and/ or ${\cal P}^-$ are possible,
subject to Eq.~(\ref{f4}) with zero right-hand side.

From Eq.~(\ref{f1}) we obtain
\begin{eqnarray}
B^2(r)=\left[1-\frac{2Gm(r)}{r}\right]^{-1}\,,
\end{eqnarray}
where the mass function is
\begin{eqnarray}
m(r)=4\pi\int^r_a\rho^{\rm eff}(r')r'^2dr'\,,
\end{eqnarray}
and $a=0$ for the interior solution, while $a=R$ for the exterior
solution. Equation~(\ref{f3}) integrates in the interior for
$\rho=$\,const to give
\begin{equation}
A^-(r)={\alpha\over \rho+p(r)}\,,
\end{equation}
where $\alpha$ is a constant.

The Israel-Darmois matching conditions at the stellar surface
$\Sigma$ give~\cite{s}
\begin{eqnarray}
[G_{\mu\nu}r^\nu]_{\Sigma}=0\,,
\end{eqnarray}
where $[f]_\Sigma\equiv f(R^+)-f(R^-)$. By the brane field
equation~(\ref{6'}), this implies $[T^{\rm
eff}_{\mu\nu}r^\nu]_\Sigma=0$, which leads to
\begin{eqnarray}  \label{m2}
\left[p^{\rm eff}+ {4\over\kappa^4\lambda}{\cal
P}\right]_\Sigma=0\,.
\end{eqnarray}
Even if the physical pressure vanishes at the surface, the
effective pressure is nonzero there, so that in general a radial
stress is needed in the exterior to balance this effective
pressure.

Assuming that the physical pressure vanishes on the surface, i.e.
$p(R)=0$, this becomes
\begin{equation}\label{m3}
(4\pi G)^2\rho^2(R)+{\cal U}^-(R) +2 {\cal P}^-(R)= {\cal U}^+(R)
+2 {\cal P}^+(R)\,.
\end{equation}
Note that we have multiplied through by $\lambda$ to obtain this
form, so that there is no general relativity limit of the
equation.

In general relativity, Eq.~(\ref{m2}) implies
\begin{equation}
p(R)=0\,,
\end{equation}
whereas for the braneworld model, we take this as a (physically
reasonable) assumption.

Equation~(\ref{m3}) gives the matching condition for any static
spherical star with vanishing pressure at the surface. If there
are no Weyl stresses in the interior, i.e. ${\cal U}^-=0= {\cal
P}^-$, and if the energy density is non-vanishing at the surface,
$\rho(R)\neq0$, then there must be Weyl stresses in the exterior,
i.e. the exterior cannot be Schwarzschild. Equivalently, {\em if
the exterior is Schwarzschild and the energy density is nonzero at
the surface, then the interior must have nonlocal Weyl stresses}.

We will further assume that ${\cal P}^-=0$, which is consistent
with the isotropy of the physical pressure in the star, so that
\begin{equation}\label{u}
{\cal U}^-(r)={\beta\over \left[A^-(r)\right]^4}\,,
\end{equation}
where $\beta$ is a constant. The matching condition in
Eq.~(\ref{m3}) then reduces for a {\em uniform} star to
\begin{equation}\label{m4}
(4\pi G)^2\rho^2+{\beta\over\alpha^4}\rho^4= {\cal U}^+(R) +2
{\cal P}^+(R)\,.
\end{equation}
It follows that the exterior of a uniform star cannot be
Schwarzschild if there are no Weyl stresses in the interior.

The Weyl stresses arise from the projection of the bulk Weyl
tensor, which responds nonlocally to the gravitational field on
the brane, and `backreacts' on the brane. Thus in general, we
expect that Weyl stresses will occur in {\em both} the interior
and exterior. However, it is possible to find consistent solutions
on the brane with Weyl stresses only in the exterior. The general
case of an interior with Weyl stresses is much more complicated.

\section {Braneworld generalization of exact uniform-density solution}

With uniform density and ${\cal U}^-=0= {\cal P}^-$, we have the
case of purely local (high-energy) modifications to the general
relativity uniform-density solution, i.e. to the Schwarzschild
interior solution~\cite{exact}. The interior mass function is
\begin{eqnarray}\label{mass}
m^-(r)=M\left[1+\frac{3M}{8\pi\lambda R^3}\right]\left({r\over R}
\right)^3\,,
\end{eqnarray}
where $M=4\pi R^3\rho/3$. Thus
\begin{eqnarray}
B^-(r)={1\over\Delta(r)}\,,
\end{eqnarray}
and the pressure is given by
\begin{eqnarray} \label{3}
{p(r)\over\rho}=\frac{[{\Delta(r)}-{\Delta(R)}] (1+\rho/\lambda)}
{[3{\Delta(R)}-{\Delta(r)}]+[3{\Delta(R)}-2{\Delta(r)}]\rho/
\lambda}\,,
\end{eqnarray}
where
\begin{eqnarray}
\Delta(r)=\left[1-{2GM\over r}\left({r\over R}\right)^3\left\{ 1+
{\rho\over 2\lambda }\right\}\right]^{1/2}\,.
\end{eqnarray}
In the general relativity limit, $\lambda^{-1}\to 0$, we regain
the known exact solution~\cite{exact}. The high-energy corrections
considerably complicate the exact solution.

Since $\Delta(R)$ must be real, we find {\em an astrophysical
lower limit on $\lambda$, independent of the Newton-law and
cosmological limits:}
\begin{equation}\label{al}
{\lambda}\geq \left(\frac{GM}{R-2GM}\right)\rho~~\mbox{for all
uniform stars}\,.
\end{equation}
In particular, this implies $R>2GM$, so that the Schwarzschild
radius is still a limiting radius, as in general relativity.
Taking a typical neutron star (assuming uniform density) with
$\rho\sim 10^{9}$~MeV$^4$ and $M\sim 4\times 10^{57}$~GeV, we find
\begin{eqnarray}\label{al'}
\lambda > 5\times 10^{8}~\mbox{MeV}^4\,.
\end{eqnarray}
This is the astrophysical limit, below which stable neutron stars
could not exist on the brane. It is much stronger than the
cosmological nucleosynthesis constraint, but much weaker than the
Newton-law lower bound. Thus stable neutron stars are easily
compatible with braneworld high-energy corrections, and the
deviations from general relativity are very small. If we used the
lower bound in Eq.~(\ref{al'}) allowed by the stellar limit, then
the corrections to general relativistic stellar models would be
significant, as illustrated in Fig.~1.

\begin{figure}
\begin{center}
\includegraphics[width=6cm,height=7cm,angle=-90]{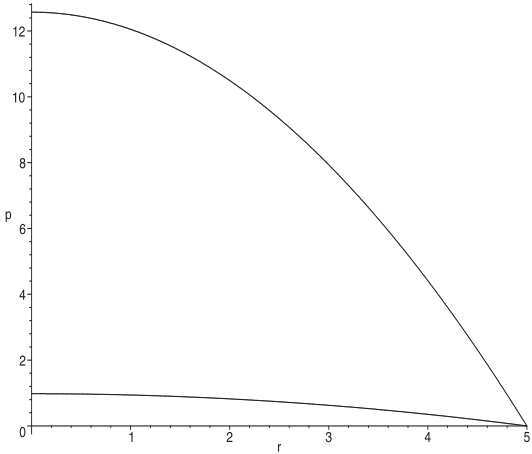}
\end{center}
\label{g1} \caption{Qualitative comparison of the pressure $p(r)$,
in general relativity (upper curve), and in a braneworld model
with $\lambda= 5\times 10^{8}$~MeV$^4$ (lower curve).}
\end{figure}

We can also obtain an upper limit on compactness from the
requirement that $p(r)$ must be finite. Since $p(r)$ is a
decreasing function, this is equivalent to the condition that
$p(0)$ is finite and positive, which gives the condition
\begin{eqnarray}\label{comp}
\frac{GM}{R} \leq {4\over 9}\left[{1+7\rho/4\lambda +
5\rho^2/8\lambda^2
\over(1+\rho/\lambda)^2(1+\rho/2\lambda)}\right]\,.
\end{eqnarray}
It follows that high-energy braneworld corrections {\em reduce}
the compactness limit of the star. For the stellar bound on
$\lambda$ given by Eq.~(\ref{al'}), the reduction would be
significant, but for the Newton-law bound, the correction to the
general relativity limit of $4\over9$ is very small. The lowest
order correction is given by
\begin{equation}\label{comp2}
{GM\over R}\leq  \frac{4}{9}\left[1-\frac{3\rho}{4\lambda}
+O\left({\rho^2\over\lambda^2}\right)\right]\,.
\end{equation}
For $\lambda\sim10^8$~GeV$^4$, the minimum allowed by
sub-millimetre experiments, and $\rho\sim 10^{9}$~MeV$^4$, the
fractional correction is $\sim 10^{-11}$.

As argued above, any exterior solution that matches this
uniform-density solution cannot be a Schwarzschild exterior. We
will now present two possible exterior solutions.

\section{Two possible non-Schwarzschild exterior solutions}

The system of equations satisfied by the exterior spacetime on the
brane is not closed. Essentially, we have two independent unknowns
${\cal U}^+$ and ${\cal P}^+$ satisfying one equation, i.e.
Eq.~(\ref{f4}) with zero right-hand side. Even requiring that the
exterior must be asymptotically Schwarzschild does not lead to a
unique solution. Further investigation of the 5-dimensional
solution is needed in order to determine what the further
constraints are. We are able to find two exterior solutions (with
${\cal U}^-=0= {\cal P}^-$) that are consistent with all equations
and matching conditions on the brane, and that are asymptotically
Schwarzschild.

The first is the Reissner-N\"ordstrom-like solution given
in~\cite{dmpr}, in which a tidal Weyl charge plays a role similar
to that of electric charge in the general relativity
Reissner-N\"ordstrom solution. We stress that there is {\em no}
electric charge in this model: nonlocal Weyl effects from the 5th
dimension lead to an energy-momentum tensor on the brane that has
the same form as that for an electric field, but without any
electric field being present. The formal similarity is not
complete, since the tidal Weyl charge gives a {\em positive}
contribution to the gravitational potential, unlike the negative
contribution of an electric charge in the general relativistic
Reissner-N\"ordstrom solution.

The braneworld solution is~\cite{dmpr}
\begin{eqnarray}
&&\left(A^+\right)^2=\left(B^+\right)^{-2}=1-\frac{2G{\cal
M}}{r}+\frac{q}{r^2}\,,\label{rn1} \\&& {\cal U}^+=-\frac{{\cal
P}^+}{2} = \frac{4}{3}\pi Gq\lambda \,{1\over r^4}\,,\label{rn2}
\end{eqnarray}
where the matching conditions imply
\begin{eqnarray}
q&=&-3GMR\,{\rho\over\lambda}\,,\\ {\cal M}&=& M \left(
1-\frac{\rho}{\lambda }\right)\,, \\ \alpha&=& \rho\Delta (R)\,.
\end{eqnarray}
Note that the Weyl energy density in the exterior is {\em
negative}, so that 5-dimensional graviton effects lead to a
strengthening of the gravitational potential (this is discussed
further in~\cite{dmpr,ssm}). Since ${\cal M}>0$ is required for
asymptotic Schwarzschild behaviour, we have a slightly stronger
condition on the brane tension:
\begin{equation}
\lambda>\rho \,.
\end{equation}
However, this still gives a weak lower limit, $\lambda>
10^{9}~\mbox{MeV}^4$. In this solution the horizon is at
\begin{equation}
r_{\rm h}=G{\cal M}\left[1+\left\{1+\left({3R\over
2GM}-2\right){\rho\over \lambda} +{\rho^2\over
\lambda^2}\right\}^{1/2}\right].
\end{equation}
Expanding this exact expression shows that the horizon is slightly
beyond the general relativistic Schwarzschild horizon:
\begin{equation}
r_{\rm h}=2GM\left[1+{3(R-2GM)\over
4GM}\,{\rho\over\lambda}\right] +O\left({\rho^2\over
\lambda^2}\right)
>2GM\,.
\end{equation}
The exterior curvature invariant ${\cal R}^2=
R_{\mu\nu}R^{\mu\nu}$ is given by
\begin{eqnarray}\label{r2}
{\cal R}= 8\pi G\left({\rho\over \lambda}\right)^2\left({R\over
r}\right)^4\,.
\end{eqnarray}
Note that for the Schwarzschild exterior, ${\cal R}=0$.

The second exterior is a new solution. Like the above solution, it
satisfies the braneworld field equations in the exterior, and the
matching conditions at the surface of the uniform-density star. It
is given by
\begin{eqnarray}
\left(A^+\right)^2&=&1-\frac{2G{\cal N}}{r}\,, \label{n1} \\
\left(B^+\right)^{-2}&=&\left(A^+\right)^2
\left[1+\frac{C}{\lambda(r-{\textstyle{3\over2}} G{\cal
N})}\right]\,,\label{n2} \\ {\cal U}^+&=& {2\pi G^2{\cal
N}C\over(1-3G{\cal N}/2r)^2}\,{1\over r^4}\,,\label{n3}\\ {\cal
P}^+&=& \left({2\over3}-{r \over G{\cal N}}\right){\cal U}^+\,.
\label{n4}
\end{eqnarray}
From the matching conditions:
\begin{eqnarray}
{\cal N} &=& M\left[\frac{1+2\rho/\lambda}{1+3GM\rho/
R\lambda)}\right]\,,\\ C &=& 3GM\rho\left[ \frac{1-3GM/2R}{ 1
+3GM\rho/R\lambda}\right]\,,\\ \alpha &=&{\rho\Delta(R)\over (1+3
GM\rho/R\lambda)^{1/2}}\,.
\end{eqnarray}
The horizon in this new solution is at
\begin{equation}
r_{\rm h}=2G{\cal N}\,,
\end{equation}
which leads to
\begin{equation}
r_{\rm h}=2GM\left[1+\left({2R-3GM\over 2R}\right){\rho\over
\lambda}\right]+O\left({\rho^2\over\lambda^2}\right)>2GM\,.
\end{equation}
The curvature invariant is
\begin{eqnarray}
{\cal R}&=&\sqrt{{\textstyle{3\over2}}}RC\left({4\pi R\over
3M}\right)^2{(1-8G{\cal N}/3r+2G^2{\cal N}^2 /r^2)^{1/2}\over
1-3G{\cal N}/2r}\times\nonumber\\ &&~~{}\times \left({\rho\over
\lambda}\right)^2\left({R\over r}\right)^3 \,.
\end{eqnarray}
Comparing with Eq.~(\ref{r2}), it is clear that these two
solutions are different. The difference in their curvature
invariants is illustrated in Fig.~2.

\begin{figure}
\begin{center}
\includegraphics[width=6cm,height=7cm,angle=-90]{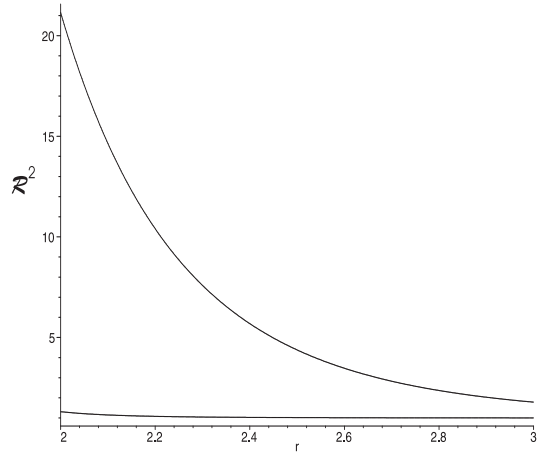}
\end{center}
\label{g3} \caption{Qualitative behavior of the curvature
invariant ${\cal R}^2$: the upper curve is the
Reissner-N\"ordstrom-like solution given by Eqs.~(\ref{rn1}) and
(\ref{rn2}); the lower curve is the new solution given by
Eqs.~(\ref{n1})--(\ref{n4}) ($\lambda= 5\times 10^{8}$~MeV$^4$).}
\end{figure}

\section{Interior solution with Schwarzschild exterior}

If we assume that the exterior is the Schwarzschild exterior
(${\cal U}^+=0={\cal P}^+$), then Eqs.~(\ref{u}) and (\ref{m4})
imply that the interior must have {\em negative} Weyl energy
density:
\begin{eqnarray}
{\cal U}^-(r)=-\left(\frac{4\pi G}{\rho}\right)^2[\rho+p(r)]^4\,.
\end{eqnarray}
This means that the tidal effects on the brane from bulk gravitons
reinforce the gravitational field in the star.
(See~\cite{dmpr,ssm} for further discussion of the meaning of
${\cal U}<0$.)

It follows that the mass function in Eq.~(\ref{mass}) becomes
\begin{eqnarray}
m^-(r)&=& M\left(1+\frac{\rho}{2\lambda
}\right) \left({r\over R} \right)^3\nonumber\\
&&~~~{}-\frac{6\pi}{\lambda\rho^2}\int^r_0 [\rho+p(r')]^4 r'^2
dr'\,,
\end{eqnarray}
which is {\em reduced} by the negative Weyl energy density,
relative to the solution in the previous section and to the
general relativity solution. The effective pressure is given by
\begin{equation}\label{pschw}
p^{\rm eff}=p-{\rho\over2\lambda}(2+6w+4w^2+w^3)\,,
\end{equation}
where $w=p/\rho$. Thus $p^{\rm eff}<p$, so that 5-dimensional
high-energy effects reduce the pressure in comparison with general
relativity.

\section{Conclusions}

We have investigated how 5-dimensional gravity can affect static
stellar solutions on the brane. We found the exact braneworld
generalization of the uniform density stellar solution, and used
this to estimate the local (high-energy) effects of bulk gravity.
We derived an astrophysical lower limit on the brane tension
$\lambda$, given by Eq.~(\ref{al}), which is much stronger than
the big bang nucleosynthesis limit, but much weaker than the
experimental Newton-law limit. We also found that the star is less
compact than in general relativity, as shown by Eqs.~(\ref{comp})
and  (\ref{comp2}). The smallness of high-energy corrections to
stellar solutions flows from the fact that $\lambda$ is well above
the energy density $\rho$ of stable stars. However nonlocal
corrections from the bulk Weyl curvature (5-dimensional graviton
effects) have qualitative implications that are very different
from general relativity.

The Schwarzschild solution is no longer the unique asymptotically
flat vacuum exterior; in general, the exterior carries an imprint
of nonlocal bulk graviton stresses. The exterior is not uniquely
determined by matching conditions on the brane, since the
5-dimensional metric is involved via the nonlocal Weyl stresses.
We demonstrated this explicitly by giving two exact exterior
solutions, both asymptotically Schwarzschild. Each exterior which
satisfies the matching conditions leads to a bulk metric, which
could in principle be determined locally by numerical integration.
However, this is very complicated even in the simpler case of
black holes on the brane~\cite{crss}. Without any exact or
approximate 5-dimensional solutions to guide us, we do not know
how the properties of the bulk metric, and in particular its
global properties, will influence the exterior solution on the
brane.

Guided by perturbative analysis of the static weak field
limit~\cite{rs,pert,ssm,dd}, we make the following conjecture:
{\em if the bulk for a static stellar solution on the brane is
asymptotically AdS$_5$ and has regular Cauchy horizon, then the
exterior vacuum which satisfies the matching conditions on the
brane is uniquely determined, and agrees with the perturbative
weak-field results at lowest order.} An immediate implication of
this conjecture is that the exterior is not Schwarzschild, since
perturbative analysis shows that there are nonzero Weyl stresses
in the exterior~\cite{ssm} (these stresses are the manifestation
on the brane of the massive Kaluza-Klein bulk graviton modes). In
addition, the two exterior solutions that we present would be
ruled out by the conjecture, since both of them violate the
perturbative result for the weak-field potential,
Eq.~(\ref{pert}).

The static problem is already complicated, so that analysis of
dynamical collapse on the brane will be very difficult. However,
the dynamical problem could give rise to more striking features.
Energy densities well above the brane tension could be reached
before horizon formation, so that high-energy corrections could be
significant. We expect that these corrections, together with the
nonlocal bulk graviton stress effects, will leave a non-static,
but transient, signature in the exterior of collapsing matter.
This is currently under investigation.

\[ \]
{\bf Acknowledgements:} We thank Edward Anderson, Bruce Bassett,
Marco Bruni, Malcolm MacCallum and Kei-ichi Maeda for useful
discussions. CG is supported by PPARC.


\begin{references}

\bibitem{rs} L. Randall and R. Sundrum, Phys. Rev. Lett. {\bf 83},
4690 (1999).

\bibitem{sms} T. Shiromizu, K. Maeda and M. Sasaki,
Phys. Rev. D {\bf 62}, 024012 (2000).

\bibitem{m2}
R. Maartens, gr-qc/0101059.

\bibitem{dmpr}
N.K. Dadhich, R. Maartens, P. Papadopoulos and V. Rezania, Phys.
Lett. {\bf B487}, 1 (2000).

\bibitem{crss}
A. Chamblin, H.S. Reall, H. Shinkai and T. Shiromizu, Phys. Rev. D
{\bf 63}, 064015 (2001).

\bibitem{pert}
J. Garriga and T. Tanaka, Phys. Rev. Lett. {\bf 84}, 2778 (2000);
C. Csaki, J. Ehrlich, T.J. Hollowood, and Y.
Shirman, Nucl. Phys. B{\bf 581}, 309 (2000); S. Giddings, E. Katz,
and L. Randall, J. High Energy Phys. {\bf 03}, 023 (2000); I.Ya.
Arafeva, M.G. Ivanov, W. Muck, K.S. Viswanathan, and I.V.
Volovich, Nucl. Phys. B{\bf 590}, 273 (2000); Z. Kakushadze, Phys.
Lett. B{\bf 497}, 125 (2000).

\bibitem{ssm}
M. Sasaki, T. Shiromizu, and K. Maeda, Phys. Rev. D {\bf 62},
024008 (2000);


\bibitem{dd}
N. Deruelle and T. Dolezel, gr-qc/0105118.

\bibitem{m}
R. Maartens, Phys. Rev. D {\bf 62}, 084023 (2000).

\bibitem{mwbh}
R. Maartens, D. Wands, B.A. Bassett and I.P.C. Heard, Phys. Rev. D
{\bf 62}, 041301 (2000).

\bibitem{s}
J.L. Synge, {\em Relativity: the General Theory} (North Holland,
Amsterdam, 1971), p39; N.O. Santos, Mon. Not. R. Ast. Soc. {\bf
216}, 403 (1985).

\bibitem{exact}
D. Kramer, H. Stephani, M. MacCallum, and E. Herlt, {\em Exact
Solutions of Einstein's Field Equations} (Cambridge, 1980).



\end{references}
\end{document}